\documentclass[fleqn,usenatbib]{mnras}

\usepackage{newtxtext,newtxmath}

\usepackage[T1]{fontenc}
\usepackage{ae,aecompl}

\usepackage{graphicx}	
\usepackage{amsmath}	
\usepackage{amssymb}	

\usepackage[normalem]{ulem}
\usepackage{booktabs}

\newcommand{\Lya}{$\rm{Ly}\alpha \,$}
\newcommand{\Ha}{$\rm{H}\alpha \,$}

\title{Are \Lya emitters segregated in protoclusters regions?}

\author[T. Hough et. al.]{
Tom\'as Hough,$^{1,2,3}$\thanks{E-mail: tomas@fcaglp.unlp.edu.ar}
Siddhartha Gurung-L\'opez,$^{4}$
 \'Alvaro Orsi,$^{5,6}$ Sof\'ia A. Cora,$^{1,2,3}$ 
 \newauthor Cedric G. Lacey,$^{7}$ and Carlton M. Baugh.$^{7}$
\vspace{0.2cm}\\
$^{1}$Instituto de Astrof\'isica de La Plata (CCT La Plata, CONICET, UNLP), Observatorio Astron\'omico,\\
Paseo del Bosque, B1900FWA La Plata, Argentina\\
$^{2}$Facultad de Ciencias Astron\'omicas y Geof\'isicas, Universidad Nacional de La Plata, Observatorio Astron\'omico,\\
Paseo del Bosque, B1900FWA La Plata, Argentina\\
$^{3}$Consejo Nacional de Investigaciones Cient\'ificas y T\'ecnicas (CONICET), Rivadavia 1917, Buenos Aires, Argentina\\
$^{4}$Department of Physics, Missouri University of Science and Technology, 1315 N. Pine Street, Rolla, MO 65409, U.S.A.\\
$^{5}$Centro de Estudios de F\'isica del Cosmos de Arag\'on, Plaza San Juan 1, piso 2, Teruel, 44001, Spain.\\
$^{6}$PlantTech Research Institute Limited. South British House, 4th Floor, 35 Grey Street, Tauranga 3110, New Zealand\\
$^{7}$Institute for Computational Cosmology, Department of Physics, University of Durham, South Road, Durham, DH1 3LE, UK.\\
}

\date{Accepted XXX. Received YYY; in original form ZZZ}

\pubyear{2020}

\begin{document}
\label{firstpage}
\pagerange{\pageref{firstpage}--\pageref{lastpage}}
\maketitle

\begin{abstract}
The presence of neutral hydrogen in the inter-stellar medium (ISM) and inter-galactic medium (IGM) induces radiative transfer (RT) effects on \Lya photons which affect the observability of \Lya emitters (LAEs).
We use the \textsc{GALFORM} semi-analytic model of galaxy formation and evolution to analyse how these effects shape the spatial distribution of LAEs with respect to \Ha emitters (HAEs) around high density regions at high redshift. We find that when a large sample of protoclusters is considered, HAEs
showing also \Lya emission (HAEs+LAEs) populate the same regions as those that do not display the \Lya line at $z=2.2$. We compare against the protocluster USS1558-003, one of the most massive protoclusters located at $z=2.53$. Our results indicate that the strong depletion of HAEs+LAEs present in the high density regions of USS1558-003 may be due to cosmic variance.
We find that at $z=2.2$ and $z=3.0$, RT of the ISM produces a strong decline ($30$-$50$ per cent) of the clustering amplitude of HAEs+LAEs with respect to HAEs towards the protoclusters centre. At $z=5.7$, given the early evolutionary state of protoclusters and galaxies, the clustering of HAEs+LAEs has a smaller variation ($10$-$20$ per cent) towards the protoclusters centre.
Depending on the equivalent width 
and luminosity criteria of the emission-line galaxy sample, the IGM can have a mild or a null effect on galaxy properties and clustering in high density regions. 

\end{abstract}

\begin{keywords}
Galaxies: high redshift --- Galaxies: clusters ---
Radiative transfer --- Intergalactic medium --- methods: numerical
\end{keywords}



\section{Introduction}
\label{sec:intro}

The properties of the large-scale environment
in which protoclusters are embedded is crucial to determine how they will evolve into massive galaxy clusters (M$_\star > 10^{14} \rm{M}_\odot$) at the present time. Protoclusters at high redshift ($z \geq 2$) are identified as overdense regions of galaxies and gas, usually associated to radio galaxies \citep[e.g. ][]{Lefevre1996, Pentericci1997, Venemans2002, Venemans2007, Hayashi2012, Orsi2016}, quasars \citep[e.g. ][]{Wold2003,Kashikawa2007, Overzier2009a, Adams2015} or other massive objects (e.g. sub-millimetre galaxies, \Lya blobs).

Emission-line galaxies (ELG) are star forming galaxies whose spectra contain intense nebular emission lines. As 
the
characteristic intensity of lines in emission 
allows detection and precise redshift, ELGs are often used to detect matter overdensities at high redshift. This helps constrain their spatial distribution over a small slice of cosmic volume.

Among ELG, those that have a detectable \Lya emision line (${\lambda=1216}$\AA) or a detectable \Ha emission line (${\lambda=6563}$\AA) are referred to as \Lya emitters (LAES) and \Ha emitters (HAEs), respectively. 
In a star-forming galaxy, these lines have the same astrophysical origin, i.e., 
they are produced
when the ionizing emission of young and massive stars is absorbed by atomic hydrogen regions located in the ISM. The recombination of these atoms leads to the emission of both \Lya and \Ha photons \citep*{Orsi2012, Dijkstra2017}. Furthermore, \Lya photons are absorbed and scattered by the inter-stellar medium (ISM), the circum-galactic medium (CGM) and the inter-galactic medium (IGM) through complex radiative transfer processes that affect the LAEs observed properties \citep{Orsi2014, Gurung2019b}. However, due to the small cross-section of the interaction between \Ha photons and neutral hydrogen atoms, HAEs are largely unaffected by these effects, making these galaxies excellent tracers of instantaneous 
star formation rate (SFR)
\citep{Kennicutt1998a, calzetti2013}.
Dual emitter surveys comprising both \Ha and \Lya emission at the same redshift are a powerful tool to understand the intrinsic properties of HAEs and LAEs, and provide insights on the intrinsic and observed \Lya luminosity functions. 
Although the fraction of dual emitters (HAEs+LAEs) depends on the survey depths \citep{Shimakawa2017b}, several studies report on the low fraction of \Lya photons that escape from \Ha emitters, from $\sim 5$ per cent \citep{Hayes2010, Matthee2016} up to $\sim 37$ per cent \citep{Sobral2017}. Escape fraction is found to strongly anti-correlate with dust extinction and SFR, and only weakly with stellar mass \citep{Hayes2011,Matthee2016}.
We remark that dual emitter surveys can be performed on a narrow redshift range ($z\sim 2.2-2.5$) from ground-based observatories.

The spatial distribution of ELGs in overdense regions, can be used to infer the underlying dark matter distribution, and how galaxy properties relate to the environment in which they reside \citep{Mo2004, Cooray2005, Overzier2006, Orsi2016, Ota2018, Shi2019, GonzalezPerez2020}.  In particular, the way in which
the environment of protoclusters affects the emission of ELG at high redshift is still matter of debate \citep[see][for a review]{Overzier2016}. Several authors have reported on different galaxy populations
residing in
high density regions. 
For instance, using data from the VLT FORS fields, 
\citet{Venemans2002}
and \citet{Venemans2007} found that LAEs are (relatively) randomly distributed in the protocluster TN-J1338, located at $z\sim 4.1$, while \citet{Overzier2008} found that LAEs seem to prefer regions devoid of UV-selected Lyman Break Galaxies.
\citet{Hayashi2016} studied the properties of HAEs that trace the rich protocluster USS1558-003, located around a radio galaxy at $z=2.53$ \citep{Hayashi2012}. They found that HAEs with $M_\star>10^{10}\rm{M_\odot}$ are located in the $\text{SFR}-M_\star$ main sequence of star-forming field galaxies, while some HAEs with $M_\star<10^{9.3}\rm{M_\odot}$ are deviated upward the main sequence, with SFRs consistent with starburst galaxies. 
\citet{Shimakawa2017a} analysed the \Lya emission of the HAEs in the  protocluster USS1558-003 and found a clear lack of LAEs in dense regions traced by HAEs, and suggested that an excess of dust and gas accreted in cold streams might prevent the escape of Ly$\alpha$ photons from the core of the protocluster. This dual emission analysis has been also performed in the SpiderWeb protocluster, located around the {\sc PKS 1138-262} radio galaxy at $z=2.16$;
this protocluster presents a concentration of \Ha emitters that increases towards the radio galaxy, while \Lya emitters do not \citep{Kurk2004}. 

A way to quantify how galaxies are distributed around a central object is computing the cross-correlation function. At high redshift, the cross-correlation function $\xi_{\rm cc}$ between overdensity tracers (radio galaxies and quasars) and ELGs can offer different information on small and large scales. \citet{Orsi2016} found that, at large scales ($r \gtrsim 10 \rm{Mpc} \, \textit{h}^{-1} $), the amplitude of $\xi_{\rm cc}$ for \Ha and \Lya emitters is
larger when the central objects are radio galaxies, 
because they inhabit more massive haloes. 
At small scales, faint LAEs ($\rm L\alpha > 10^{41} erg\, s^{-1}\, \it{h}^{-2}$) have higher $\xi_{\rm cc}$ than bright LAEs ($\rm L\alpha > 10^{42} erg\, s^{-1}\, \it{h}^{-2}$), because AGN feedback prevent starburst galaxies to dominate the galaxy abundance at small separations.
Recently, \citet{GurungLopez_2020} found that the presence of the IGM induces a scale-dependent effect on the auto-correlation function $\xi(\rm r)$ of LAEs at $z=5.7$, where the shape of $\xi(\rm r)$ becomes broader at the baryon acoustic oscillation scale, and the maximum is displaced by ${\sim 1 \, \rm{cMpc} \, \textit{h}^{-1} }$. 
If the presence of the IGM had an impact on the observed spatial distribution of LAEs in high density environments, it could produce misleading conclusions on the interpretation of clustering data of future surveys such as HETDEX \citep{hill2008}, DESI \citep{levi2013}. 
 
In this work we use the GALFORM semi-analytic model of galaxy formation and evolution \citep{Cole2000, Lacey2016, Baugh2019}  to explore the spatial distribution of 
HAEs and LAEs around protoclusters, at redshifts up to $z\lesssim 6$, and evaluate the impact of the IGM in such distributions. 

In Section ~\ref{sec:tools}, we describe the semi-analytic model and dark matter only simulation on which the model is applied, along with a brief description of our theoretical approach of the radiative transfer process that takes place in both the ISM and IGM.
In Section ~\ref{sec:neighbor}, we analyse the spatial distribution of LAEs and HAEs in high density environments. The impact of the IGM on the clustering of LAEs at small scales around protoclusters is detailed in Section ~\ref{sec:clustering}. Our conclusions are summarized in Section ~\ref{sec:conclusions}.


\section{Theoretical approach}
\label{sec:tools}
The construction of a realistic synthetic population of galaxies in a cosmological context requires a set of numerical tools that combines the cosmological framework with baryonic physics, which rules the intrinsic and observable properties of galaxies. This is achieved by combining a cosmological dark matter simulation with a semi-analytic model of galaxy formation, and open-source software that incorporate ISM and IGM radiative transfer effects. 

\begin{itemize}
    
    \item \textbf{Dark matter only simulation.} The P-Millennium \citep{Baugh2019} is a state-of-the-art dark matter only \textit{N}-body simulation 
    that models the hierarchical growth of structures in the $\Lambda \rm{CDM}$ scenario. It
    uses the Planck cosmology: $\rm{H}_0 = 67.77~\rm{km\, s^{-1}\,Mpc^{-1}}$, $\Omega_\Lambda = 0.693$, $\Omega_{\rm{M}} = 0.307$ , $\sigma_8 = 0.8288$ \citep{Planck2016}.
    The box size 
    is $542.16~ {\rm cMpc}\, h^{-1}$ 
    and the particle mass ${M_{\rm p} = 1.061 \times 10^8~{\rm M}_\odot\, h^{-1}}$
    ($5040^3$ dark matter particles). 
    The dark matter halo merger trees are constructed from the SUBFIND subhaloes using the DHALOS algorithm described in \citet{Jiang2014}. Haloes 
    that contain at least 20 particles are retained, corresponding to a halo mass resolution limit of $2.12 \times 10^9~{\rm M}_\odot \, h^{-1} $. \\

    \item \textbf{Semi-analytic model.} We use the $\rm GALFORM$ semi-analytic model of galaxy formation and evolution.
    In short, $\rm GALFORM$ initially populates dark matter haloes with gas. Then, tracking the merger history of haloes, the gas is evolved  including several physical mechanisms: 
    i) shock-heating and radiative cooling of gas inside haloes; ii) formation of a galactic disk with quiescent star formation; iii) 
    triggering of starburst episodes 
    in bulges due to disk instabilities and mergers; 
    iv) active galactic nuclei,
    supernovae and photoionization feedback to regulate the star formation rate; v) the chemical evolution of gas and stars. \\

    $\rm GALFORM$ computes the \Ha and \Lya luminosities of galaxies from the total production rate of hydrogen ionizing photons (Lyman continuum photons). This is obtained by integrating the composite spectral energy distribution (SED) of each galaxy over the extreme-UV continuum down to the Lyman break at ${\lambda = 912 \mathring{A}}$. Then, by assuming that all of these ionising photons are absorbed within the ISM of the galaxy and that no direct recombination into ground state takes place (case B recombination), a fraction of Lyman continuum photons is converted into different line fluxes \citep{osterbrock1989,Dijkstra2014a}. 
    On one hand, $\rm H\alpha$ emission can suffer dust attenuation.
    $\rm GALFORM$ includes a two-step dust attenuation: one for the emission of stars that are still inside their birth cloud, and one for the emission that emerges from molecular clouds and stars located outside the clouds, which are affected by diffuse dust component present in the disk/bulge components of the galaxy. This model includes diffuse dust attenuation at 14 bands, including the R band (centered at $6594 \mathring{A}$). We refer the reader to Appendix A of \citet{Lacey2016} for more details.
    On the other hand, the intrinsic luminosity of \Lya photons is expected to be reduced by both the scattering they suffer by neutral hydrogen atoms in the ISM and IGM, and their absorption by dust grains. \\

    \item \textbf{ISM radiative transfer model.} \Lya photons are assumed to escape the galaxy through outflows. 
    The outflow is characterized by an expansion velocity, hydrogen column density and dust optical depth, which depend upon the galaxy properties. 
    The outflow velocity is computed as:
    
    \begin{equation}
        \rm{V_{exp,c}} = \kappa_{V,c}{\rm SFR_c}\frac{r_c}{M_\star}
    \end{equation}
    
    \noindent where the subscript $c$ denotes the galaxy component (disk or bulge), ${\rm SFR}_c$ is the star formation rate in $M_\odot {\rm Gyr} h^{-1}$, $r_c$ is the half stellar mass radius in ${\rm Mpc} h^{-1}$, $M_\star$ is the total stellar mass of the galaxy in ${\rm M}_\odot h^{-1}$, and $\kappa_{V,c}$ are free dimensionless parameters that regualate the efficiency of gas ejection.
    The neutral hydrogen column density of the outflows is computed for each component as:
    
    \begin{equation}
        N_{\rm H,c} = \kappa_{\rm N,c}\frac{M_{\rm cold,c}}{r_{\rm c}^2}
        \label{eq:density}
    \end{equation}
    
    \noindent where $M_{\rm cold,c}$ is the cold gas mass of each component in ${\rm M}_\odot h^{-1}$ units, and $\kappa_{\rm N,c}$ are free parameters calibrated for each component and redshift.
    Finally, the optical depth of dust absorption is computed as:
    
    \begin{equation}
        \tau_{\rm a,c} = (1-\rm{A}_{\rm Ly_\alpha})\frac{E_\odot}{Z_\odot}N_{\rm H,c}Z_{\rm c}
    \end{equation}
    
    \noindent where $E=1.77\times 10^{-21}\rm{cm}^{-2}$ is the ratio $\tau_{\rm a}/N_{\rm H}$ for solar metallicity, $\rm A_{\rm Ly\alpha}$ is the albedo at $\rm Ly\alpha$ wavelenght, the solar metallicity is $Z_\odot=0.02$ and $Z_{\rm c}$ is the metallicity of the cold gas of each component. \\

    Then, in order to compute the escape fraction ($f_{\rm esc}$), we use FLaREON \citep{GurungLopez_2019b}, an open Python package based on a Monte Carlo RT code  \citep{Orsi2012} that predicts the \Lya line profiles and escape fractions of photons in outflows of different characteristics.
    The FLaREON code includes three different gas outflow geometries: Thin Shell, Galactic Wind (both with spherical symmetry but different neutral hydrogen density profiles) and Bicone
    \citep[see][for further details]{GurungLopez_2019b}. In this work, we use the Thin Shell geometry to compute the ISM transmission, where the hydrogen column density of the outflow is described by Eq.~\ref{eq:density}. The Thin Shell geometry
    reproduces better the observed properties of \Lya emitters, like the dependence of the offset of the peak of the \Lya line on stellar mass, SFR and $\rm EW$ (Gurung-Lopez et al. 2020b, in preparation).\\

    \item \textbf{IGM radiative transfer model.} While inside galaxies the losses of \Lya flux are due to dust absorption, photons in the IGM are scattered out of the line of sight by the neutral hydrogen. 

    We estimate the IGM transmission for every galaxy depending on the local environmental properties, such as density, velocity and ionization state of the IGM. In the simulation, the IGM is distributed in cosmological boxes of $\sim 2 {\rm cMpc}h^{-1}$ a side, with its density determined according to the DM content inside the box.

    As a first approximation, the IGM absorbs photons with wavelengths shorter than $1216$\AA.
    Moreover, as galaxies lie in overdense regions, the IGM opacity is higher close to the galaxy, causing the drop in the transmission close to \Lya wavelength. Then, the IGM transmission flattens to the IGM cosmic transmission. Additionally,
    the mean number density of neutral hydrogen atoms in the IGM increases with redshift up to $z \lesssim 6$, leading to an increase of the optical depth as well \citep[see][for a review]{McQuinn2016}. The average transmission of the IGM results of $85 \%$ ($40 \%$) for $z=2$ ($z=4$) \citep{Dijkstra2014a}, and drops below $1$ per cent at $z=5.7$ for higher frequencies than that of \Lya \citep{GurungLopez_2019b}.

\end{itemize}


\section{LAE depletion at high densities} \label{sec:neighbor}

\begin{figure*}
    \centering
    \includegraphics{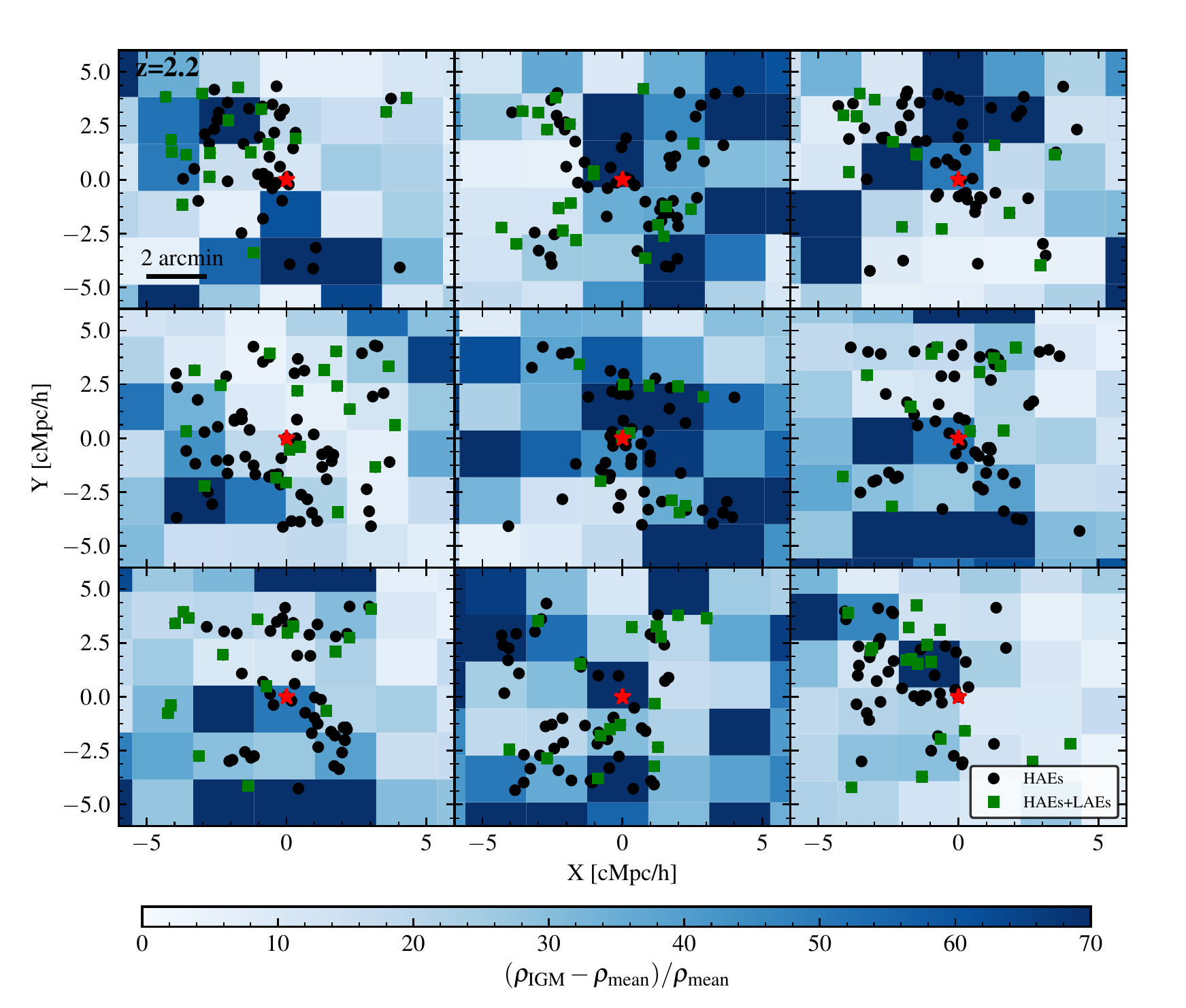}
    \caption{Spatial distribution of HAEs (black dots) and HAEs+LAEs (green squares) in 9 simulated protoclusters at ${z=2.2}$. Positions are given in co-movil coordinates.
    Each protocluster has $M_{\rm halo}>10^{13.7} {\rm M_\odot} h^{-1}$ (the red star indicates the central radio galaxy). The spatial constraints, $\rm EW$ and luminosity criteria are adopted from S17 (FL sample). The coloured squares represent the maximum value of the IGM overdensity (considering the extension of the protocluster along the $z$ coordinate) with respect to the mean density of IGM in the complete simulation.
    We highlight that the 3 protoclusters of the upper panel
    are those that
    present a HAEs+LAEs depletion similar to what is present in USS1558-003 protocluster, located at ${z=2.53}$ (S17).
    }
    \label{fig:spatial}
\end{figure*}


\citet{Shimakawa2017a} (S17 hereafter) studied the \Lya emission of HAEs located in USS1558-003, the richest protocluster known at $z \sim 2.5$, with an estimated dynamical mass of $\sim 10^{14} \rm{M_\odot}$, consistent with a progenitor of a massive cluster ($\sim 10^{15} \rm{M_\odot}$) in the local Universe \citep{Shimakawa2014}. They found that LAEs tend to avoid high density regions traced by HAEs, and that denser regions present lower \Lya escape fractions. This could be produced by a gaseous and dusty component covering the protocluster core, and it is not clear whether this should be expected as systematic for other protoclusters, or is the result of a particular conjunction of intrinsic characteristics. 

Motivated by these observational results, 
we use $\rm GALFORM$ semi-analytic model
to study the spatial segregation of LAEs relative to HAEs in a wide sample of simulated protoclusters. 
As in S17, we study the relation between the spatial distribution of galaxies and the local density of galaxies that exhibit \Ha and \Lya emission simultaneously (HAEs+LAEs). 
In their work, S17 performed \Lya imaging using the $\rm NB428$ narrow band filter (central wavelength of $4297$ \AA~and FWHM of $84$ \AA) of the Subaru Prime Focus Camera of the Subaru Telescope. The FWHM that was used allows the detection of LAEs
with ${z = 2.53 \pm 0.03}$, or ${21.3 \, \rm Mpc}$ uncertainty depth. S17 included the observations in \Ha performed by \citet{Hayashi2016}, and the ELG sample consisted of 104 HAEs, with 13 of those galaxies also presenting \Lya emission.

We create mock catalogues of protoclusters at ${z=2.2}$ with the same spatial constraints as the one observed by S17. 
To do this, 
from all our central galaxies, we consider radio-galaxy candidates as a protocluster centre at high redshift.  Radio galaxies were selected according to the halo mass function \citep{Orsi2016}. At $z=2.2$, we have $1048$ protoclusters candidates with $M_{\rm halo}> 10^{13.2}~ {\rm M}_\odot \,h^{-1}$. The mean density of galaxies inside $2~\rm cMpc$ of these objects spans between $10$ and $400$ times the mean density of objects in our simulation.

The distance to the $N{\rm th}$
neighbour is commonly used as a proxy for local density, and has the advantage of not assuming an underlying geometry \citep{Baldry2006, Bluck2019}. In the case of S17, they define the mean projected distance
${\rm <a>}_{N\rm{th}}=2 \times (\pi \sum _{N {\rm th}})^{-0.5}$, where $\sum _{N{\rm th}} (= N / \pi r_{N{\rm th}}^2)$ is the number density of galaxies within the $r_{N{\rm th}}$ radius. This is the distance to the $(N-1)\rm{th}$ neighbour from each galaxy, and $N=5$.

By applying emission-line equivalent width, $\rm EW$, and luminosity limits, we define two samples of simulated ELG around protoclusters at $z=2.2$:

\begin{itemize}
    \item{ Flux limited sample (referred to as FL)}: we consider the same EW and luminosity limits as
    in S17 
    \citep[see also][for more details]{Hayashi2016, Shimakawa2017b}. In this case, HAEs have line emission widths $\rm EW_{H\alpha}>18.6~\mathring{\rm A}$ and luminosity $L_{\rm{H}\alpha}>4.35 \times 10^{41}~ \rm{erg\, s^{-1}}$, while HAEs+LAEs are imposed to have also $\rm EW_{Ly\alpha}>15~\mathring{\rm A}$ and $L_{{\rm Ly}\alpha}>4.4 \times 10^{41}~ \rm{erg\,s^{-1}}$ (i.e., HAEs+LAEs satisfy both luminosity and $\rm EWs$ cuts).
    
    \item{ Fixed number density sample (referred to as FN)}: we impose luminosity limits that allow us to match the observed surface density of HAEs and HAEs+LAEs as in S17. In this case, HAEs have $L_{\rm{H}\alpha}>10^{41}~ \rm{erg\, s^{-1}}$, while HAEs+LAEs are imposed to have also $L_{{\rm Ly}\alpha}>1.5 \times 10^{42}~ \rm{erg\,s^{-1}}$. The $\rm EW$ limits are equivalent as in the first sample. With these limits, our protocluster candidates have a median of $90$ HAEs and $13$ HAEs+LAEs. 
\end{itemize}


\begin{figure} 
\centering
\includegraphics[width=\columnwidth]{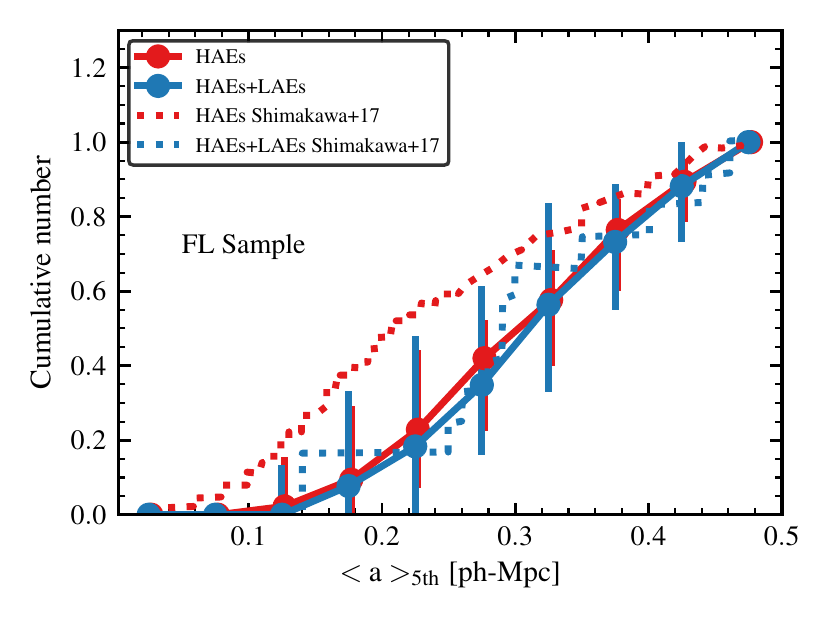}
\includegraphics[width=\columnwidth]{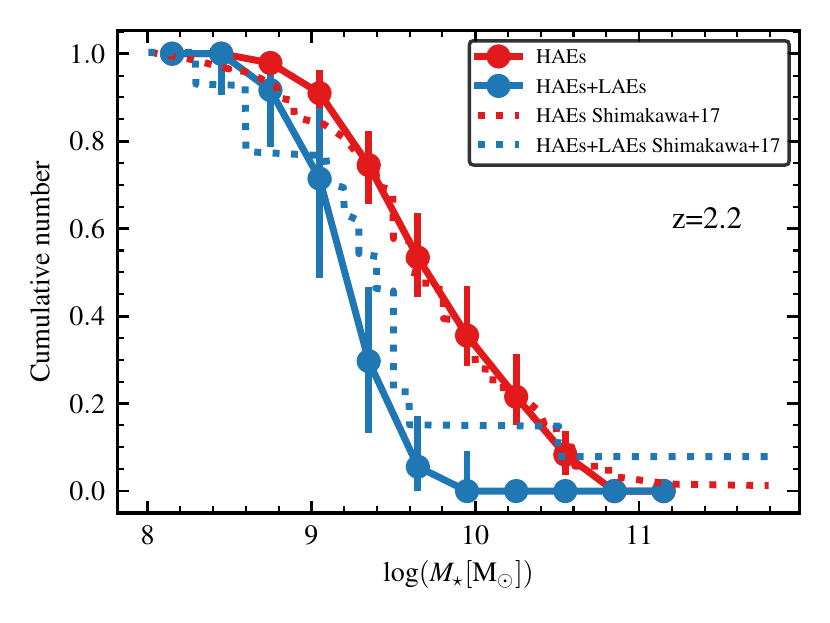}

\caption{Cumulative number of HAE and HAEs+LAE in terms of the mean projected distance $<\rm a>_{5th}$ (upper panel) and stellar mass content (lower panel) for the FL sample. Solid lines represent the median and error bars denote the $10-90th$ percentiles for
HAEs (red) and HAEs+LAEs (blue) for the 30 most massive protoclusters selected at $z=2.2$, who present halo masses above $10^{13.7} \rm{M_\odot}\, \textit{h}^{-1}$. Dotted lines represent the behaviour of the USS1558-003 protocluster, located at $z=2.53$. The selection of synthetic HAEs and HAES+LAES
matches the observational conditions of \citet{Shimakawa2017a}. 
}
\label{fig:cumulative1}
\end{figure}

\begin{figure}
\centering
\includegraphics[width=\columnwidth]{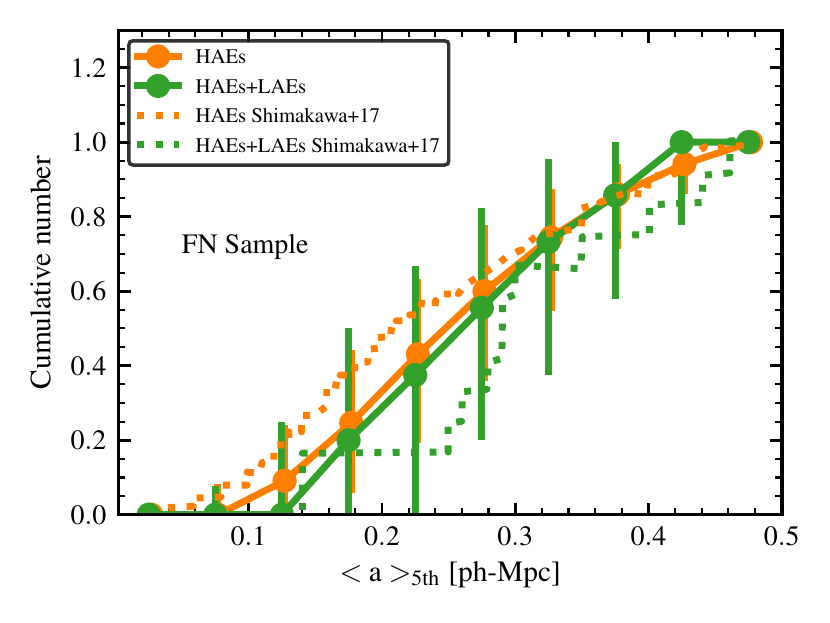}
\includegraphics[width=\columnwidth]{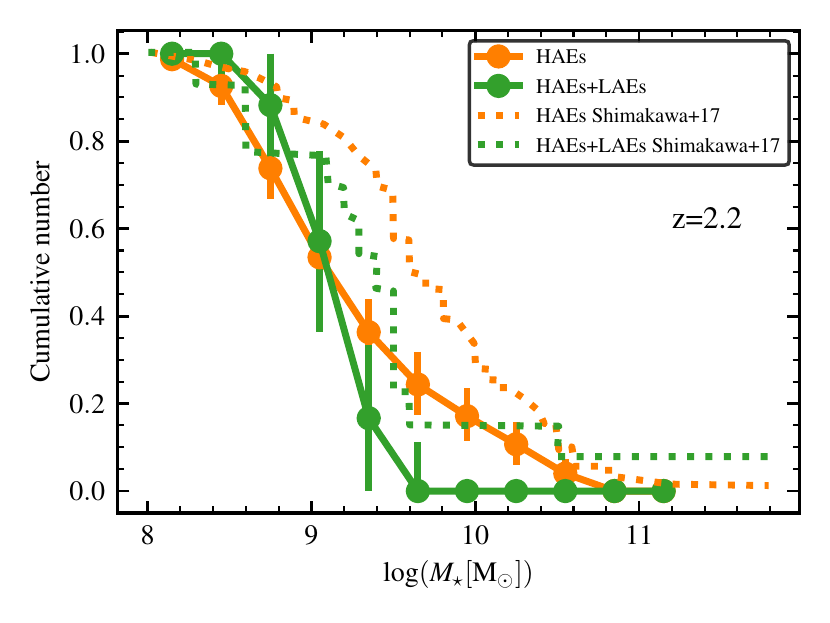}

\caption{Same as Fig.~\ref{fig:cumulative1}, but for the FN sample. Solid lines represent the median and error bars denote the $10-90th$ percentiles of HAEs (orange) and HAEs+LAEs (green) for 1048 protoclusters selected at $z=2.2$, who present halo masses above $10^{13.2} \rm{M_\odot}\, \textit{h}^{-1}$. Dotted lines represent the behaviour of the USS1558-003 protocluster. The selection of HAEs and HAEs+LAEs was defined to match the surface density of ELGs in USS1558-003 \citep{Shimakawa2017a}.
}
\label{fig:cumulative2}
\end{figure}

\begin{figure}
    \centering
    \includegraphics{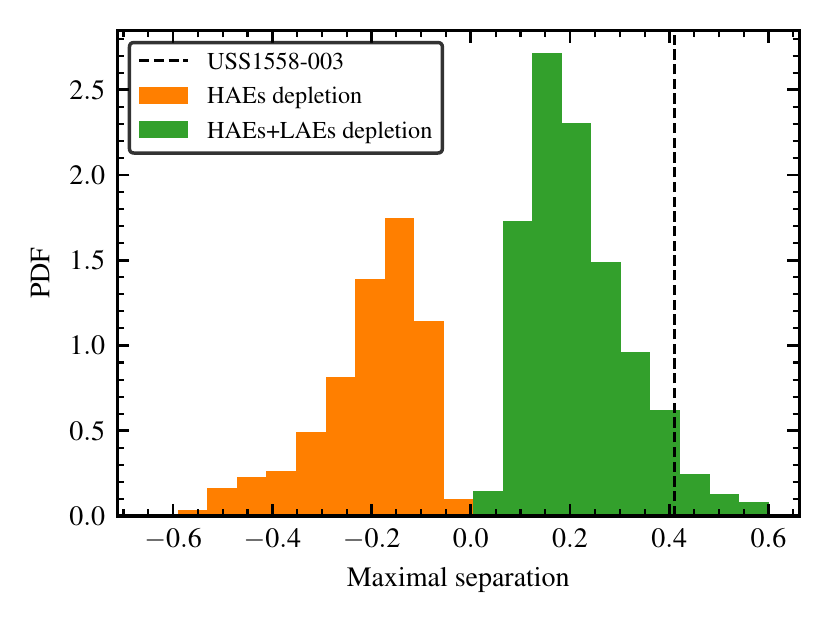}
    \caption{Histogram of the K-S tests between cumulative numbers of HAEs and HAEs+LAEs for all simulated protocluster considered in Fig.~\ref{fig:cumulative2} (FN sample). Positive values (in green) are associated to HAEs+LAEs depletion (as in S17), while negative values (in orange) are associated with HAEs depletion. Near $\sim 10$ per cent of the simulated protoclusters have distances consistent with USS1558-003, represented by the black vertical line.
    }
    \label{fig:histo}
\end{figure}


For both samples, ELGs located inside a $2.0\times3.5\times21.5 ~\rm Mpc$ box centered on each protocluster centre 
are considered members of the sample. These spatial constraints correspond to those of S17. 

Although the protoclusters 
in both our FN and FL samples span
halo masses between $10^{13.2}\rm{M_\odot}\, \textit{h}^{-1}$ and $10^{14.2} \rm{M_\odot}\, \textit{h}^{-1}$
(a mass range that comprises the value of the dynamical mass of the protocluster USS1558-003),
none of the protoclusters in the FL sample reaches the number density of HAEs in USS1558-003.

Reproducing high density environments at high redshift is challenging for simulations, mainly due to failures in capturing correctly the baryonic physics that are involved in the assembly of stellar mass with cosmic time.
To approximate this high density environment, we use only the 30 most massive protoclusters of our FL sample,
with halo masses above $10^{13.7} \rm{M_\odot}\, \textit{h}^{-1}$
and an average of $45$ HAEs and $10$ HAEs+LAEs.
The fact that the number density of HAEs in these protoclusters do not reach the number density of USS1558-003 could induce a bias in our analysis: a higher number of HAEs would certainly decrease ${\rm <a>_{\rm 5th}}$, affecting the spatial distribution of both populations. However, we consider that the environment in which galaxies reside is well characterized, as the simulated dark matter halo masses encompasses the inferred $M_{\rm halo}$ of USS1558-003, and we highlight that the analysis of both FN and FL samples points towards the same conclusion.

In Fig.~\ref{fig:spatial} we show the spatial distribution of HAEs and HAEs+LAEs in a subsample of the 30 most massive protoclusters of the FL sample. In colored squares we represent the maximal IGM overdensity (along the line of sight, in the volume ocuppied by the protocluster) with respect to the median IGM density in the complete simulation. 
The protoclusters of the upper row
are those that
present a depletion of HAEs+LAEs in high local densities (${\rm <a>_{\rm 5th}}<0.3$), similar to what is found in S17, as we discuss in Fig.~\ref{fig:cumulative1}.
While HAEs in most simulated protoclusters seem to distribute around the central radio galaxy, in some cases the distribution have a clear offset of $\sim 2$ arcmin, similar to S17. The middle and right panels of the bottom row are examples of such protoclusters.
We note the presence of a
correlation between IGM and HAEs overdensities, despite of the
spatial fluctuations of the IGM density. As can be appreciated, the IGM model consists of homogeneous cubes of $\sim 2$ cMpc a side, where the IGM density is associated with the DM content of the box; this might result as a rough approximation. However, given the spatial constraints imposed by the observations, each simulated protocluster 
comprises
$\sim 640$ IGM cubes and, 
for each galaxy, the transmission of the \Lya line is computed through the cubes along the line of sight.

A more realistic IGM model would need a more refined grid, which would produce prohibitively expensive computational time due to the large size of the DM simulation box. 

For the FL sample,
we compute the mean projected distance $\rm{<a>_{5th}}$ for every HAE and compare the median cumulative number of HAEs and HAEs+LAEs in the upper panel of Fig.~\ref{fig:cumulative1}. Error bars represent $10$-$90th$ percentiles.
We find a clear discrepancy with observations: while S17 find that HAEs+LAEs avoid the densest regions, our analysis indicate that, on average, HAEs+LAEs inhabit the same regions as HAEs. We notice that some individual protoclusters present depletion of LAEs, while others are depleted of HAEs in high density regions. This results in similar median behaviour, but with a relatively high dispersion when the 30 protoclusters are considered. 

In general, galaxies that inhabit dense environments tend to be more massive than field galaxies, at low and intermediate redshifts \citep{Baldry2006, Darvish2015}. At $z\gtrsim2$, HAEs in protoclusters have been found to be more massive than HAEs in the field. We 
note that our simulated HAEs present an increase in stellar mass towards the protocluster centre: HAEs present between $1.3$ and $3$ times the stellar mass of HAEs located in average regions, in consistency with observations \citep{Hatch2011b, Koyama2013}. 
In the lower panel of Fig.~\ref{fig:cumulative1}, we present the stellar mass distribution of both populations of the FL sample, showing a remarkable agreement with the observational data
given by S17. \Lya radiative transfer favours the escape of \Lya photons from galaxies with lower stellar mass, dust content and SFRs than HAEs with the same line luminosity \citep{Guaita2011, Orsi2016, Shimakawa2017b}, and is further detailed in Sec.~\ref{sec:clustering}. Consistently, our HAEs+LAEs have lower stellar masses than HAEs, and start to accumulate at ${\rm log}(M_\star/ {\rm M}_\odot) \sim 9.6$, while HAEs start to accumulate at $\rm{ log}(M_\star/ \rm{M_\odot}) \sim 10.8$.
This means that although the FL sample has lower number density of HAEs than USS1558-003, the sample 
is thus complete as the stellar masses of both HAEs and HAEs+LAEs are in accordance with observations.

In order to explore environments with similar ELG number density than USS1558-003, we perform the same analysis for the FN sample, which
considers
the complete 1048 protocluster candidates at $z=2.2$.
In this case, for a given value of the mean local density $\rm <a>_{5th}$, the number density of HAEs increases with respect to the one obtained with the FL sample, achieving a better agreement with S17, as shown in the upper panel of Fig.~\ref{fig:cumulative2}.
Nevertheless, HAEs+LAEs
do not seem to specifically avoid the regions traced by HAEs in the FN sample either. 
In fact, some protoclusters follow
the observational trend, while others present the opposite behaviour, as 
can be appreciated from
Fig.~\ref{fig:histo}.
To quantify the depletion of HAEs or LAEs in individual protoclusters, we compute a two-sample KS test between the cumulative distributions of HAEs and HAEs+LAEs. 
In Fig.~\ref{fig:histo} we show which separations are more likely to occur in  the FN sample. We assign positive values for protoclusters which present HAEs+LAEs depletion, and negative values for protoclusters that show HAEs depletion.
It is clear that a small depletion of HAEs+LAEs is the most likely scenario ($62$ per cent present HAE+LAE depletion). But a similar depletion of HAEs is also found, which results in a statistically negligible depletion of HAEs+LAEs with respect to HAEs when all protoclusters are averaged.

The lower panel of Fig.~\ref{fig:cumulative2} shows that both HAEs and HAEs+LAEs 
in the FN sample
have lower stellar masses than 
observed. This is due to the fact that in the FN sample we are considering HAEs with lower luminosity than in the FL sample, so their stellar masses are also lower. Besides, although HAEs start to accumulate at higher stellar mass than HAEs+LAEs, this trend is not as steep as observed.

As suggested by S17, the accretion of cold streams that supply the protocluster core with HI gas could prevent \Lya photons from escaping from the dense regions of USS1558-003. We find that when a large sample of protoclusters is considered, this characteristic signature should not be expected to be as violent as in S17, although a small segregation of LAEs is the most likely scenario. Probably, the result found by S17 corresponds to a particular situation. If the inflow of cold gas occurs along a stream in the line of sight, it could enhance the dispersion of \Lya photons and diminish the number of HAEs+LAEs detected in dense regions. In Fig.~\ref{fig:histo}, the dotted vertical line represents the maximum separation in USS1558-003. 
Among the simulated protoclusters 
that present HAEs+LAEs depletion, we find that $\sim 10$ per cent follow the observational trend of S17, suggesting that their result is subject to cosmic variance. However, only in $\sim 1$ per cent of the protoclusters the null hypothesis can be rejected at a $95$ per cent confidence level.

It is noteworthy that the results obtained with the FN sample are
independent of the halo mass limit of our protocluster candidates, as we find the same general behaviour when selecting haloes with ${\rm log}(M_{\rm halo} [{\rm M}_\odot \, h^{-1}])>13.5$ and ${\rm log}(M_{\rm halo}[{\rm M}_\odot \, h^{-1}])>14.0$. 
Moreover, the protoclusters that present \Lya depletion consistent with S17 show no specific signature in
their intrinsic properties, such as $\rm sSFR$, metallicity or dark matter content, with respect to those protoclusters that do not present \Lya depletion, or even with those that present \Ha depletion. This analysis 
sustains
the hypothesis that a specific environmental effect could be producing the observed \Lya depletion.

Due to the powerful multi-wavelength emission of AGNs, it is expected a certain degree of AGN contamination in NB observational samples of ELGs. For instance, \citet{Sobral2016} studied a sample of $59$ high luminous HAEs ($\rm L_{\rm H\alpha}>10^{42} ~erg s^{-1} $) 
in the redshift range
$0.8<z<2.23$, and found that $\sim 30$ per cent of the galaxies hosts AGNs, but the AGN fraction increases with \Ha luminosity, and has little to no dependence on redshift. In a sample of $188$ \Lya emitters located at $z=2.23$, \citet{Sobral2017} showed that only for $\rm L>10^{43}~ erg \,s^{-1}$ the luminosity function is dominated by X-ray AGNs. In our model, both \Ha and \Lya luminosities are powered only by the star-forming regions inside galaxies, i.e., our HAEs and HAEs+LAEs samples are not contaminated with AGNs.


\begin{table*}
    \centering
    \begin{tabular}{c | c c c | c c c} 
         & \multicolumn{3}{c|}{FL Sample [$\times 10^{-3}\, \rm{cMpc^{-3}\textit{h}^{-3}}$]}  & \multicolumn{3}{c|}{FN Sample [$\times 10^{-3}\, \rm{cMpc^{-3}\textit{h}^{-3}}$]} \\
        \cmidrule(lr){2-4}\cmidrule(lr){5-7}
         & HAEs & HAEs+LAEs noIGM & HAEs+LAEs IGM & HAEs & HAEs+LAEs noIGM & HAEs+LAEs IGM \\
        \hline
         \textbf{z=2.2} &  19.9 & 4.97 & 4.95 & 60.4 & 9.09 & 8.07 \\
         \textbf{z=3.0} &  26.7 & 7.79 & 7.68 & 76.9 & 13.3 & 10.3 \\
         \textbf{z=5.7} &  15.8 & 7.85 & 7.81 & 45.5 & 17.6 & 14.8 \\
         
         \bottomrule
    \end{tabular}

    \caption{Number density of ELG around protoclusters, when using different $\rm EW$, \Ha and \Lya luminosity limits. In the \textbf{FL Sample}, HAEs have $L_{\rm{H}\alpha}>4.35 \times 10^{41}~ \rm{erg\, s^{-1}}$, while HAEs+LAEs are imposed to have also $L_{{\rm Ly}\alpha}>4.4 \times 10^{41}~ \rm{erg\,s^{-1}}$. In the \textbf{FN Sample}, HAEs have $L_{\rm{H}\alpha}>10^{41}~ \rm{erg\, s^{-1}}$, while HAEs+LAEs are imposed to have also $L_{{\rm Ly}\alpha}>1.5 \times 10^{42}~ \rm{erg\,s^{-1}}$. In both cases, HAEs have $\rm EW>18~\mathring{A}$ and LAEs have $\rm EW>15~\mathring{A}$.
    The number density is computed as the number of galaxies inside  $10~ {\rm cMpc}\, h^{-1}$ of all protoclusters, divided by the volume of all protoclusters at each redshift.
    }
    \label{tab:density}
\end{table*}

\begin{figure*}
    \centering
    \includegraphics[width=6.9in]{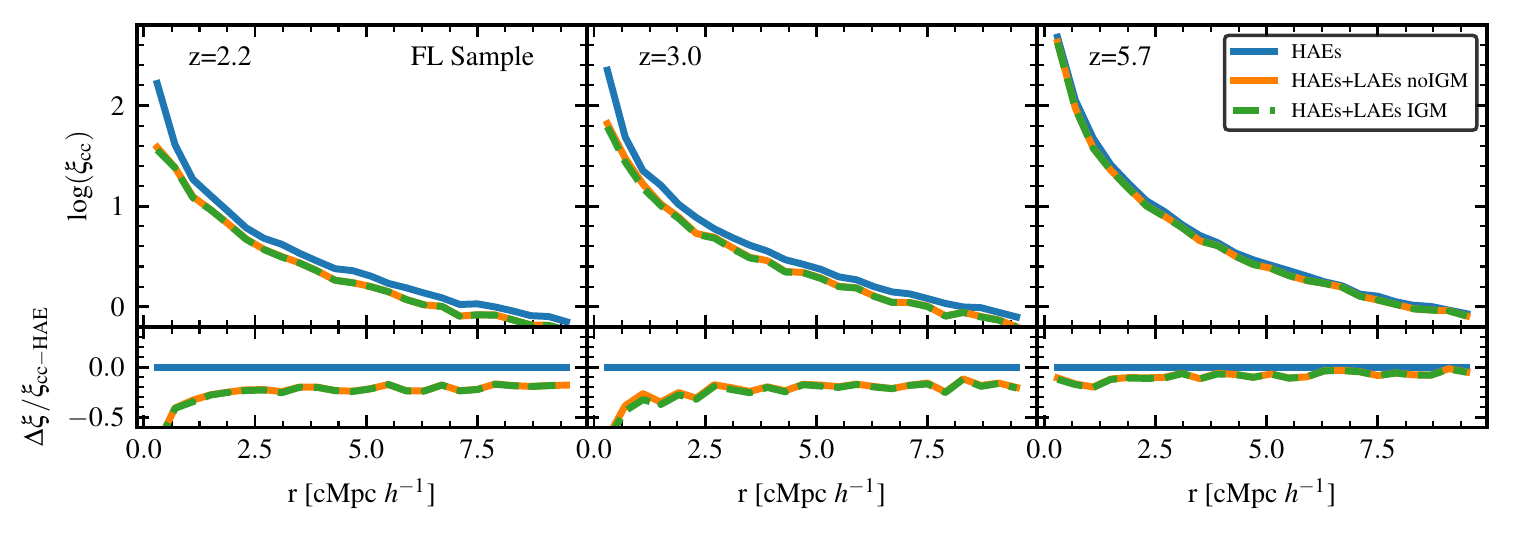}
    \caption{Cross-correlation functions for HAEs and HAEs+LAEs for ${z=2.2}$ (left panel), ${z=3.0}$ (middle panel) and ${z=5.7}$ (right panel). The galaxies are classified as HAEs or LAEs following \citet{Shimakawa2017a}, where HAEs have $\rm EW>18~\mathring{A}$ and $L_{\rm H\alpha} > 4.35 \times 10^{41}\, \rm{erg s^{-1}}$ and HAEs+LAEs are imposed to have also $L_{\rm Ly\alpha} > 4.4 \times 10^{41}\, \rm{erg\,s^{-1}}$ and $\rm EW>15~\mathring{A}$ (FL Sample defined in Sec.~\ref{sec:neighbor}). 
    Solid blue lines represent HAEs, which are not affected by IGM at any redshift. Dashed green and solid orange lines represent HAEs+LAEs from the model with and without IGM effect, respectively.
}
    \label{fig:clustering}
\end{figure*}

\begin{figure*}
    \centering
    \includegraphics[width=6.9in]{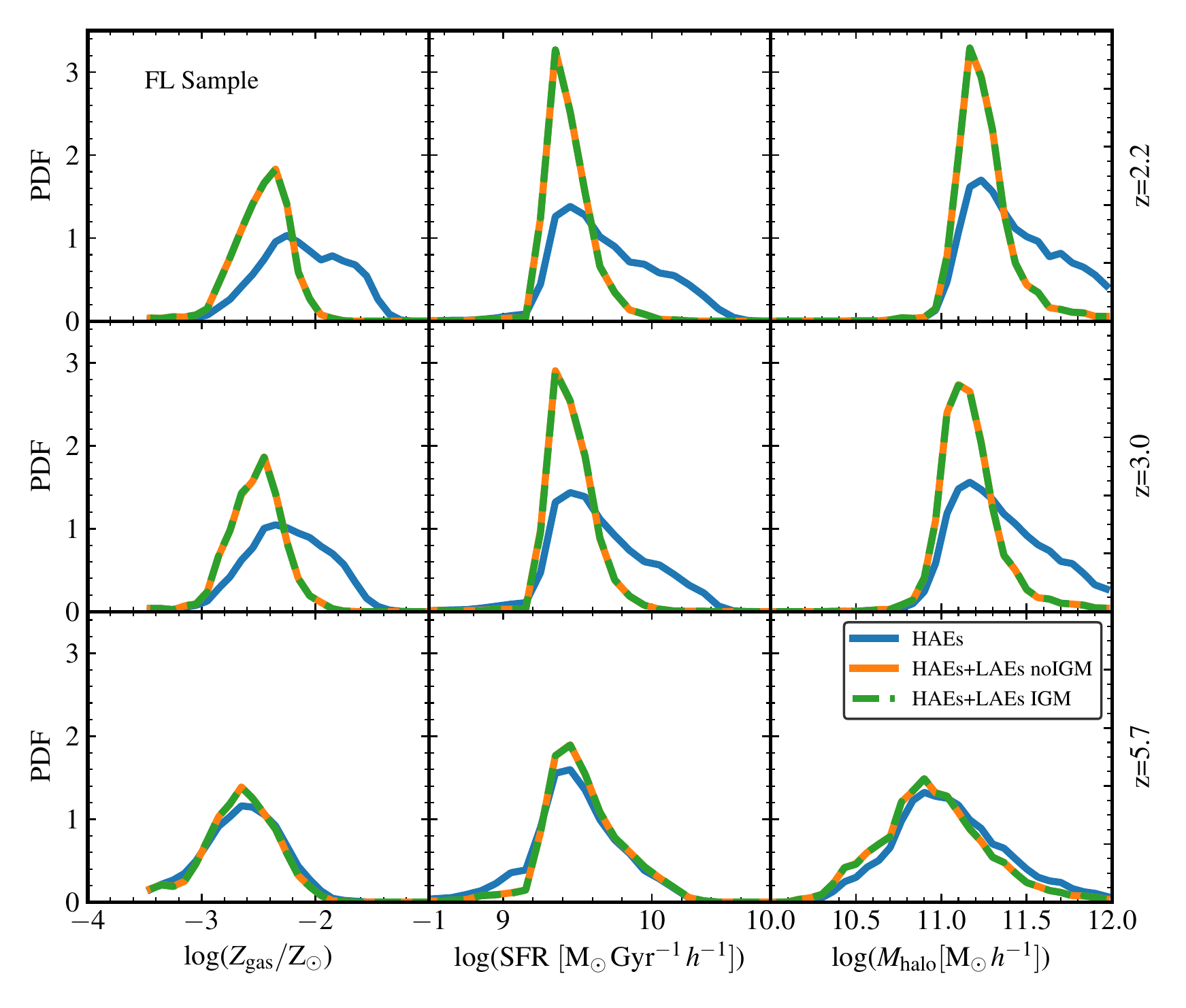}
    
    \caption{Cold gas metallicity (left column), star-formation rate (middle column) and mass of dark matter halo (right column) for ELGs
    located at distances of $\rm r < 10 \, cMpc\, \textit{h}^{-1}$ from protoclusters centre. Protoclusters are selected according to their dark matter halo mass, at three different redshifts: $z=2.2$ (upper panels), $z=3.0$ (middle panels) and $z=5.7$ (lower panels). In each case, HAEs have $\rm EW>18~\mathring{A}$ and $L_{\rm H\alpha} > 43.5 \times 10^{41}\, \rm{erg\, s^{-1}}$ and HAEs+LAEs are imposed to have also $L_{\rm Ly\alpha} > 4.4 \times 10^{41}\, \rm{erg\,s^{-1}}$ and $\rm EW>15~\mathring{A}$ (as in the FL Sample of Sec.~\ref{sec:neighbor}). The properties of HAEs+LAEs are the same for the simulation with and without IGM effect included, hence the presence of the IGM has a negligible impact on the HAEs+LAEs population of the FL sample at all redshifts. 
    }
    \label{fig:properties}
\end{figure*}

\begin{figure*}
    \centering
    \includegraphics[width=6.9in]{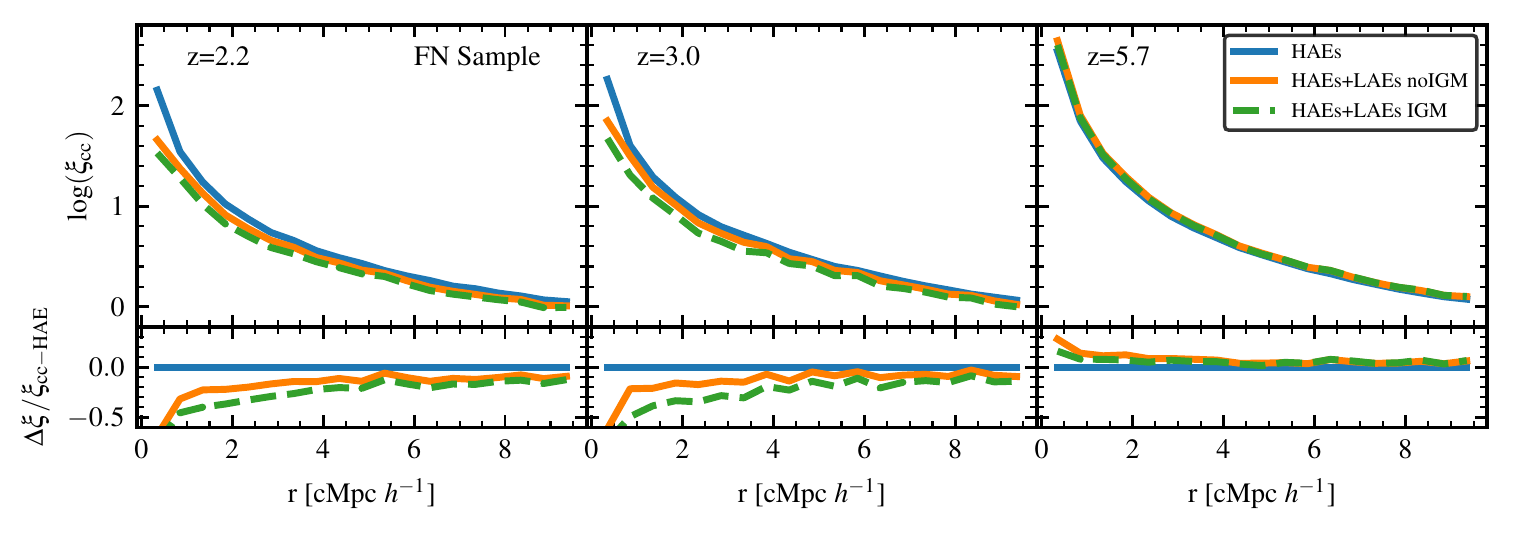}
    \caption{Same as Fig.~\ref{fig:clustering}. In this case, HAEs have $\rm EW>18~\mathring{A}$ and $L_{\rm H\alpha} > 10^{41}\, \rm{erg\,s^{-1}}$ and HAEs+LAEs are imposed to have as $L_{\rm Ly\alpha} > 1.5 \times 10^{42}\, \rm{erg\,s^{-1}}$ and $\rm EW>15~\mathring{A}$ (as in 
    FN Sample of Sec.~\ref{sec:neighbor}).
    }
    \label{fig:clustering2}
\end{figure*}

\begin{figure*}
    \centering
    \includegraphics[width=6.9in]{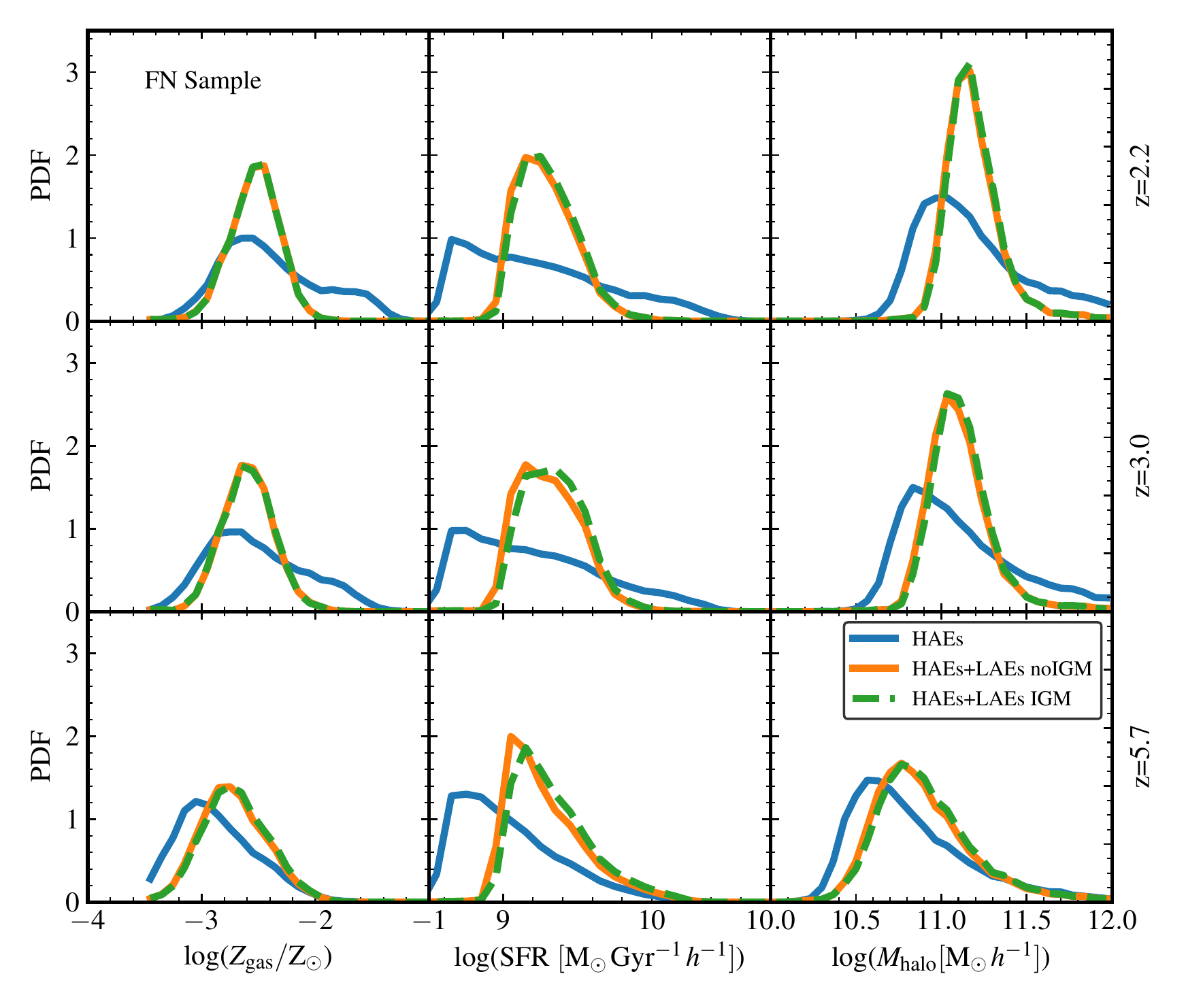}
    \caption{Same as Fig.~\ref{fig:properties}. In this case, HAEs have $\rm EW>18~\mathring{A}$ and $L_{\rm H\alpha} > 10^{41}\, \rm{erg\,s^{-1}}$ and HAEs+LAEs are imposed to have as $L_{\rm Ly\alpha} > 1.5 \times 10^{42}\, \rm{erg\,s^{-1}}$ and $\rm EW>15~\mathring{A}$ (as in 
    FN Sample of Sec.~\ref{sec:neighbor}). 
    }
    \label{fig:properties2}
\end{figure*}


\section{The impact of the IGM on the clustering at small scales}
\label{sec:clustering}

In the $\Lambda \rm{CDM}$ paradigm, the density of the IGM is higher around massive structures, increasing the probability of 
scattering
$\rm{Ly}\alpha$ photons that escape from star-forming galaxies. 
\citet{GurungLopez_2020} used the semi-analytic model and radiative transfer tools detailed in Sec.~\ref{sec:tools} to explore the coupling between the IGM and the \Lya observability, and  
found that the IGM modifies the clustering amplitude of LAEs on scales $> 20\, {\rm cMpc}\,h^{-1}$, while at scales $<5\,{\rm cMpc} \, h^{-1}$ the clustering of LAEs seems unaffected by the presence of the IGM.
Although \citet{GurungLopez_2020} studied the clustering of LAEs in a wide variety of scales, they did not take into account the local environment in which galaxies reside. The impact of IGM on the clustering of LAEs at small scales might differ in more extreme environments. 

In this section, we explore the coupling between IGM and HAEs+LAEs in high density environments by comparing the clustering of HAEs and HAEs+LAEs at small scales ($<10\,{\rm cMpc} \, h^{-1}$) for the model with and without IGM included (IGM and noIGM model, respectively). We select central galaxies as protoclusters centres according to the host halo mass distribution at ${z=2.2}$, ${z=3.0}$, and ${z=5.7}$. The selection results in $1048$, $579$ and $1564$ candidates with halo masses of $ M_{\rm halo}>10^{13.2} \rm{M_\odot} $, $ M_{\rm halo}>10^{13} \rm{M_\odot} $ and $ M_{\rm halo}>10^{12} \rm{M_\odot} $ for ${z=2.2}$, ${z=3.0}$ and ${z=5.7}$, respectively. These limits correspond to the peak of the host halo mass distribution of radio galaxies \citep{Orsi2016}. The limit for halo mass of central galaxies at $z=5.7$ is somewhat arbitrary, but our findings are insensitive to these specific values; 
we arrive to the same conclusions when repeating our analysis varying the limits in  $0.2$ dex.

It is worth mentioning
that at ${z=2.2}$, the protoclusters candidates are equivalent to those of Sec.~\ref{sec:neighbor}. 
We select HAEs and HAEs+LAEs inside a spherical shell of radius
$r = 10\,{ \rm cMpc} \, h^{-1}$ 
from the central object. 
To define the ELGs samples at different redshifts, we maintain the same luminosity and $\rm EW$ criteria that define FL and FN samples in Sec.~\ref{sec:neighbor}. We caution the reader that, in this section, the spatial constraints to select ELGs are modified with respect to Sec.~\ref{sec:neighbor}, thus the samples at ${z=2.2}$ do not match exactly. However, the properties of both samples discussed further in this section are insensitive to the different spatial constraints aforementioned. Hence, we will keep the sample
nomenclature defined in Sec.~\ref{sec:neighbor}, specifying the redshift considered when necessary.

The number density of objects is computed as the sum over all protoclusters
of HAEs and HAEs+LAEs with protoclustercentric distances less than $10~ {\rm cMpc}\,h^{-1}$, divided by the sum of the volumes of the protoclusters. The values for both samples at all redshifts considered are listed in table \ref{tab:density}.

We quantify the clustering as the cross-correlation function between halo mass selected central objects and ELGs, $\xi_{\rm cc}$. This is estimated as
\begin{equation}
    \xi_{\rm cc} (r) = \frac{DD(r)}{N_{\rm c} n_{\rm gal}\Delta V(r)} -1,
    \label{eq:xi}
\end{equation}
\noindent where $ DD(r)$ is the total number of ELGs around central objects at a distance $r \pm \Delta r/2$, $N_{\rm{c}}$ is the total number of protocluster candidates at the corresponding redshift, $n_{\rm gal}$ is the mean number density of ELGs in the box, and $\Delta V(r)$ is the volume of a spherical shell of radius $r$ and width $\Delta r$. This width corresponds to the bin size used to compute $\xi_{\rm cc}$. 
As our simulation box is periodic, the pair counts are not affected by edge effects, so the use of estimators that rely on random set of objects is not necessary.\\

In Fig.~\ref{fig:clustering}, we show the clustering of the 
HAEs and HAEs+LAEs in the FL sample 
at three different redshifts, and compute the relative difference with respect to the clustering of the HAEs population as $\Delta \xi = (\xi_{\rm cc}-\xi_{\rm cc-HAE})/\xi_{\rm cc-HAE}$. 
We find that for $z=2.2$ and $z=3.0$, HAEs+LAEs are $\sim 50$ per cent less clustered than HAEs in the core of the protoclusters, and $\sim 20$ per cent less clustered 
from 
$r \sim 2.5~{\rm cMpc}\, h^{-1}$ to the outskirts of the protocluster (left and middle panel of Fig.~\ref{fig:clustering}). 
At $z=5.7$, the clustering of HAEs+LAEs is $\sim 15$-$20$ per cent smaller in the core, and 
$\lesssim 10$
per cent for $r\gtrsim 2~{\rm cMpc}\, h^{-1}$. In all cases, the results from the IGM model are basically indistinguishable from those obtained from
the noIGM model: given the combination of \Lya and \Ha luminosity thresholds, the vast majority of HAEs+LAEs in the noIGM model are also classified as HAEs+LAEs in the IGM model, hence their clustering 
are almost identical. 

In Fig.~\ref{fig:properties}, we analyse 
the metal content of the cold gas phase, $Z_{\rm gas}$, $\rm SFR$ and $M_{\rm halo}$ of ELGs located at distances of ${\rm r < 10~ cMpc} \,h^{-1}$ from the protoclusters centers at ${z=2.2}$, ${z=3.0}$ and ${z=5.7}$ for the FL sample. 
Observations suggest that \Lya emission is strongly dependent on the 
stellar mass, $M_\star$, of galaxies. Massive star-forming systems have higher gas mass content \citep{Keres2005} (which increases the scattering of \Lya photons), and have been forming stars longer, which leads to higher metal and dust content in the ISM. In fact, the anti-correlation between $f_{\rm esc}$  of \Lya photons with $M_\star$ and $\rm SFR$ has been reported at ${z\sim 2}$ \citep{Matthee2016} and $3<z<4.6$ \citep{Oyarzun2017}. 

In our case, both HAEs and HAEs+LAEs have low $Z_{\rm gas}$ at ${z=5.7}$, and their probability density function (PDF) are
very similar. This allows the escape of \Lya photons from both populations and promotes a selection of galaxies with similar $\rm SFR$ (lower left and middle panels of Fig.~\ref{fig:properties}). HAEs+LAEs tend to inhabit haloes slightly less massive than HAEs,
hence their clustering is smaller in the central and outer regions of the protoclusters, as it was already mentioned (right panel of Fig.~\ref{fig:clustering}). The chemical enrichment of the ISM due to stellar evolution increases the metal content within galaxies with cosmic time, and
thus, at lower redshifts, 
\Lya photons are more likely to escape from galaxies with intermediate metallicities ($-3<{\rm log(Z_{gas}/Z_\odot)}<-2$) and intermediate SFRs (${\rm 9.0<log(SFR \, [M_\odot Gyr^{-1} } \, h^{-1}])<9.8$), as we show respectively
in the upper and middle panels of the left and middle
columns in Fig.~\ref{fig:properties}.
In our model, HAEs+LAEs with intermediate $Z_{\rm gas}$ and intermediate $\rm SFR$ statistically inhabit DM haloes clearly less massive ($10.9<{\rm log}(M_{\rm halo} [\rm{M}_\odot])<11.5$) than HAEs ($10.9<{\rm log}(M_{\rm halo} [\rm{M}_\odot])<12.0$), both at $z=2.2$ and $z=3.0$. Hence, the clustering amplitude with respect to HAEs is lower. 

We conclude that for the FL sample, radiative transfer processes that occur inside HAEs+LAEs raise a selection effect over galactic properties, which results on a strong decrease on the clustering amplitude with respect to the HAEs population at ${z=2.2}$ and ${z=3.0}$.
We emphasize that IGM produces no enhancement of this effect, not
even at high redshift.

In the case of the FN sample, when we lower the \Ha luminosity limit for HAEs and raise the \Lya luminosity 
limit for HAEs+LAEs with respect to the FL Sample, we find that the IGM effect becomes noticeable. The number density of HAEs+LAEs in the IGM model results lower than in the noIGM model, as can be noted in the second column of Tab.~\ref{tab:density}. In this case, a small difference arises in the clustering of HAEs+LAEs between both models (Fig.~\ref{fig:clustering2}). At $z=2.2$ and $z=3.0$, HAEs+LAEs are less clustered than HAEs, as in the FL sample. For $r \leq 4\, \rm cMpc\, \it{h}^{-1}$, the IGM diminishes the clustering of HAEs+LAEs, and the slope is also less pronounced than in the FL sample.

At $z=2.2$ and $z=3.0$, HAEs still have higher gas metal content than HAEs+LAEs, as the FL sample (left column of Fig.~\ref{fig:properties2}). HAEs have $-3 \lesssim {\rm log(}Z_{\rm gas}) \lesssim -1.5$, while HAEs+LAEs have $-3 \lesssim {\rm log(}Z_{\rm gas}) \lesssim -2$, and no dependence with the IGM is noticed. But the low \Ha luminosity threshold results in HAEs with low $\rm SFR$, and the peak of the PDF is approximately $10^{8.7} {\rm M_\odot Gyr^{-1}}\, h^{-1}$ (middle column of Fig.~\ref{fig:properties2}), while the peak of the PDF for HAEs+LAEs is approximately $10^{9.3} {\rm M_\odot Gyr^{-1}}\, h^{-1}$. However, HAEs can reach $\rm SFR \sim 10^{10.4}\,M_\odot Gyr^{-1} \, h^{-1}
$, $0.5$ dex higher than HAEs+LAEs, which have values restricted to $10^{9} \lesssim 
{\rm SFR} \, [{\rm M}_\odot {\rm Gyr}^{-1} \, h^{-1}]
\lesssim 10^{9.9}$. The presence of the IGM only slightly decreases the $\rm SFR$ of HAEs+LAEs.

A similar behaviour is found for the mass of the DM haloes. The peak of the PDF for HAEs is approximately $10^{10.9} {\rm M_\odot} \, h^{-1}$, while for HAEs+LAEs it is approximately $10^{11.1} {\rm M_\odot} \, h^{-1}$ (right column of Fig.~\ref{fig:properties2}). However, HAEs
can inhabit more massive haloes than HAEs+LAEs, as they can reach values up to $10^{12} {\rm M}_\odot \, h^{-1}$, while HAEs+LAEs reach up to $10^{  11.5}{\rm M}_\odot \,h^{-1}$.
More massive DM haloes are expected to locate towards the centre of the protocluster \citep{Orsi2016}, hence those galaxies in the FN sample that inhabit the most massive haloes and have higher SFRs dominate the clustering behaviour.

At ${z=5.7}$, the FN sample have HAEs+LAEs that present higher
metal content, higher $\rm SFR$ and inhabit more massive DM haloes than HAEs (lower panels of Fig.~\ref{fig:properties2}),
hence the clustering of HAEs+LAEs results to be $\sim 30$-$10$ per cent higher for  $r<2  \, {\rm cMpc} \,{h}^{-1}$ with respect to HAEs, for the model without IGM included. When IGM is considered, the clustering of HAEs+LAEs results in  $20$-$10$ per cent higher than HAEs for  $r<2  \, {\rm cMpc} \, {h}^{-1}$. For $r>2.5  \, {\rm cMpc} \, {h}^{-1}$, the clustering of HAEs and HAEs+LAEs is very similar for both models. 

Fig.~\ref{fig:properties} and Fig.~\ref{fig:properties2} allow us to conclude that the radiative processes that take place inside galaxies shape the observable properties of ELGs, while IGM has only a minor impact.
The IGM density (computed as described in Fig.~\ref{fig:spatial}) between $8$ and $10$ cMpc$h^{-1}$ from the protocluster centre spans between $\sim 5$ and $\sim 12$ times the mean IGM density of the simulation, reflecting that protoclusters are embedded in an extensive and over-dense matter distribution. However, if we restrict to ELGs located at $r<5$ cMpc$h^{-1}$, we still do not appreciate a substantial difference in the properties for the simulations with and without IGM.
The effect of the IGM on the \Lya transmission depends on the density of the IGM in which the ELGs reside. In particular, \citet{GurungLopez_2020} ranked their LAE samples according to the IGM density where they reside, and split them into $3$ sub-samples: under-dense (below the percentile $33$ of density), intermediate (between the $33$ and $66$ percentiles) and over-dense (above the $66$ percentile). They show that at $z=2.2$ ($z=3.0$), the transmission for wavelengths bluer than \Lya ($\lambda \sim 1214$ \AA) is $0.9$ ($0.4$), $0.85$ ($0.2$) and $0.8$ ($0.1$) for under-dense, intermediate and over-dense environments, respectively. 
For $z=5.7$, the transmission remains below $1$ per cent even in under-dense regions. This means that IGM is attenuating LAEs specifically in over-dense regions at $z=2.2$ and $z=3.0$, while at $z=5.7$ the effect is the same throughout all environments. When this is combined with the permissive \Lya luminosity limit of the FL sample, the vast mayority of HAEs+LAEs remain in the sample when the IGM is included. In the case of the FN sample, 
the low \Ha luminosity limit results in higher number densities (with respect to the FL sample) for HAEs and HAEs+LAEs in both models, as can be appreciated in Tab.~\ref{tab:density}, but given the restrictive \Lya luminosity limit of the FN sample, a higher proportion of HAEs+LAEs are excluded when the IGM is included. 

From the results of our models, we conclude that the clustering of ELGs in high density environments is clearly dominated by the radiative transfer processes inside galaxies, and the IGM plays a secondary role in decreasing the clustering, even at high redshift. 
A shallow and wide survey, targeting bright LAEs, is more prone to detect galaxies affected by the IGM rather than a deep small survey.


\section{Discussion and Conclusions}
\label{sec:conclusions}
We study the possible spatial segregation of LAEs with respect to HAEs around a wide sample of protoclusters, following the observational work performed by \citet{Shimakawa2017a} (S17) on the protocluster USS1558-003, located at $z=2.53$. With this aim, we create 
catalogues of ELGs that include
Ly$\alpha$ radiative transfer of both the ISM and IGM, by combining a cosmological dark matter simulation (P-Millennium) with a semi-analytic model of galaxy formation (GALFORM).
We define two samples of HAEs and HAEs+LAEs at ${z=2.2}$. On the one hand, we built
a sample designed to reproduce the same constraints imposed by the observational work (FL sample). In this sample, HAEs have line emission widths $\rm EW_{H\alpha}>18.6~\mathring{\rm A}$ and luminosity $L_{\rm{H}\alpha}>4.35 \times 10^{41}~ \rm{erg\, s^{-1}}$, while HAEs+LAEs are imposed to have also $\rm EW_{Ly\alpha}>15~\mathring{\rm A}$ and $L_{{\rm Ly}\alpha}>4.4 \times 10^{41}~ \rm{erg\,s^{-1}}$. On the other hand, we consider
a sample designed to reproduce the observed surface density of HAEs and HAEs+LAEs (FN sample). In this sample, HAEs have $L_{\rm{H}\alpha}> 10^{41}~ \rm{erg\, s^{-1}}$, while HAEs+LAEs have also $L_{{\rm Ly}\alpha}>1.5 \times 10^{42}~ \rm{erg\,s^{-1}}$, and we maintain the same $\rm EW$ cut as for the FL Sample. We also explore how the radiative transfer of the IGM affects the clustering of FL and FN samples at $z=2.2$, $z=3.0$ and $z=5.7$, by comparing models with and without IGM radiative transfer effect included. Our main results are summarized as follows:

\begin{itemize}
    \item We average the behaviour of simulated protoclusters at $z=2.2$ for both FL and FN samples, and do not find the high depletion of HAEs+LAEs in the densest regions of protoclusters present in USS1558-003. 
    
    \item Only $\sim 10$ per cent of the simulated protoclusters are in consistency with the high HAEs+LAEs depletion present in USS1558-003, suggesting that the observational result could be subject to cosmic variance. 
    
    \item We analyse the clustering of 
    ELGs in the FL and FN samples up to ${\rm 10~ cMpc} \, h^{-1}$ from protoclusters centre. We find that radiative transfer processes inside galaxies create selection effects over galaxy properties for both samples. For the FL sample, HAEs+LAEs tend to have lower SFRs, lower metallicities and inhabit less massive haloes than HAEs at ${z=2.2}$ and ${z=3.0}$.
    
    \item  In the FL sample, the clustering of HAEs+LAEs turns out to be ${\sim 50}$ per cent lower than that of HAEs in the core of protoclusters ($r<1~ {\rm cMpc}\,h^{-1}$), and ${\sim 20}$ per cent lower in the outskirts ($r>2.5~ {\rm cMpc}\,h^{-1}$). For ${z=5.7}$, the clustering of HAEs+LAEs is between $\sim 10$-$20$ per cent smaller than HAEs in the protocluster core, and less than $10$ per cent smaller for $r > 2~{\rm cMpc} \,h^{-1}$.
    
    \item The properties of HAEs+LAEs in the FL sample are not affected by the presence of the IGM in the model, hence the clustering of both models are almost
    identical. 
    This indicates that in a survey with the capability to detect HAEs and LAEs with $\rm EW\gtrsim 15~\mathring{\rm A}$ and $L\gtrsim 4.4 \times 10^{41}~ \rm{erg\,s^{-1}}$ at $z\leq 5.7$ for both \Ha and \Lya, the clustering of HAEs+LAEs should not be affected by the presence of the IGM.
    
    \item In the case of the FN sample, the low \Ha luminosity threshold allows the inclusion of HAEs with lower SFRs and less massive DM haloes than the FL sample. Nevertheless, near the protocluster centre, the clustering is dominated by massive DM haloes, which tend to be HAEs hosts. The clustering of HAEs+LAEs is ${\sim 40}$ per cent lower than HAEs in the core of protoclusters, and ${\sim 15}$ per cent lower for $r > 4~ {\rm cMpc} h^{-1}$ at ${z=2.2}$ and ${z=3.0}$. 
    
    The presence of the IGM results in a lower clustering amplitude and less pronounced slope at $ r < 4~ {\rm cMpc} \,h^{-1}$ with respect to the noIGM model.
    
    \item Due to the restrictive \Lya luminosity threshold of the FN sample, at ${z=5.7}$, HAEs+LAEs tend to have higher SFRs and metallicities, and inhabit more massive DM haloes than pure HAEs, although the difference is small. This results in a $\sim 5$-$30$ per cent higher clustering amplitude for HAEs+LAEs at $r<2~ {\rm cMpc} \, h^{-1}$. When the IGM is included, the clustering varies only from $\sim 5$ to $15$ per cent in the central region. For $r>2~ {\rm cMpc} \, h^{-1}$ the clustering of HAEs and HAEs+LAEs is very similar.
    
\end{itemize}

S17 suggest that the accretion of cold gas streams
directly into the core of the protocluster could prevent \Lya photons from escaping, resulting in a lack of LAEs in high density regions. 
Although gas accretion along the line of sight could enhance the depletion, we claim that, on average, HAEs and HAEs+LAEs can trace similar local densities at ${z=2.2}$.
Moreover, S17 suggest that, 
as mean projected distances are
small in high density regions, Lya photons that escape from HAEs may penetrate the CGM of a foreground galaxy, increasing the depletion. Although our model does not include radiative transfer of an extended CGM component around each galaxy, we expect this effect to be small. 
Observations suggest that ionized, dense ($n > 1 \,{\rm cm^{-3}}$), and relatively cold ($\rm{T} \sim 10^4~ \rm{K}$) reservoirs of gas should surround  massive galaxies at $z \geq 2$ \citep{cantalupo2017}. In some cases, such as radio-loud and radio-quiet quasars, the Ly$\alpha$ emission from these gaseous haloes can be traced out to $100 ~\rm kpc$ of galactic radii. However, in faint LAEs (with surface brightness $\rm SB \gtrsim 4\times10^{21}~ erg s^{-1} cm^{-2} arcsec^{-2}$) \Lya haloes can reach $\sim 60~ \rm kpc$ of galactic radii \citep[see][and references therein]{wisotzki2018,witstok2019}. The properties of the CGM around high redshift galaxies are still matter of debate, and the results of our model are not sensible to the effect that CGM might produce on background LAEs. 
Moreover, the resolution of our IGM model presumes a limitation to our analysis. A reduction of the grid by a factor of $4$ would be ideal to reach the scale of the local density $<\rm a>_{\rm 5th}$ at which S17 finds the depletion of HAEs+LAEs, thus allowing to confirm our results. 
Multi narrow-band surveys such as J-PAS \citep{benitez2014} will provide a large ELGs sample at ${z \sim 2}$, where our model suggests that LAEs can inhabit high densities traced by HAEs.\\

The properties and clustering of HAEs+LAEs residing in high density regions depend mainly on the radiative transfer effects that happen inside them. IGM presence results in a second order effect, which
depends on the $\rm EW$ and luminosity criteria that defines the ELG sample.
The next generation of multi-wavelength imaging surveys will be able to characterise high redshift environments of over-dense regions with unprecedented detail. In particular, a number of these will rely on emission-line galaxies to map matter distribution. Spectroscopic surveys such as HETDEX \citep{hill2008} and DESI \citep{levi2013} have sufficient spectral resolution to probe the scales on which our model predicts that 
radiative transfer of the ISM induces a selection effect on ELGs properties, 
giving rise to a different clustering amplitude.

\section*{Acknowledgements}
We thank the referee for his/her useful comments, that 
helped improve this manuscript.
TH acknowledge {\it Consejo Nacional de Investigaciones
Cient\'{\i}ficas y T\'ecnicas} (CONICET), Argentina, for their supporting fellowships.
AO acknowledges support from project AYA2015-66211-C2-2  of  the Spanish {\it Ministerio de Econom\'ia, Industria y Competitividad.}
SAC acknowledges funding from CONICET (PIP-0387), 
and {\it Universidad Nacional de La Plata} (G11-150),
Argentina.
This project has received funding from the European Union's Horizon 2020 Research and Innovation Programme under the Marie Sklodowska-Curie grant agreement No 734374. Project acronym: LACEGAL.
The programs and figures had been developed under \textsc{Python} language \citep{Python3}, using specially \textsc{NumPy} \citep{Van2011numpy} and \textsc{Matlotlib} \citep{Hunter2007}.

\section*{Data Availability}
The data underlying this article will be shared on reasonable request to the corresponding author.




\bibliographystyle{mnras}
\bibliography{biblio_ELG}



\bsp	
\label{lastpage}
\end{document}